\documentclass[a4paper,fleqn,usenatbib]{mnras}

\usepackage{txfonts}
\usepackage{wasysym}

\usepackage[T1]{fontenc}
\usepackage{ae,aecompl}

\usepackage{psfig}   
\usepackage{graphicx}
\usepackage{amssymb}
\usepackage{subfigure}

\title[Was Planet~9 captured?]{Was Planet~9 captured in the Sun's natal star-forming region?}

\author[R.~J.~Parker, T.~Lichtenberg \& S.~P.~Quanz]{
  Richard J.~Parker$^1$\thanks{E-mail: R.Parker@sheffield.ac.uk}\thanks{Royal Society Dorothy Hodgkin Fellow}, Tim Lichtenberg$^{2,3}$ and Sascha P. Quanz$^{3}$\thanks{National Center of Competence in Research ``PlanetS'' (http://nccr-planets.ch)}
  \vspace*{0.1cm}\\
  $^1$ Department of Physics and Astronomy, The University of Sheffield, Hicks Building, Hounsfield Road, Sheffield, S3 7RH, UK \\
  $^2$ Institute of Geophysics, ETH Z{\"u}rich, Sonneggstrasse 5, CH-8092 Z{\"u}rich, Switzerland\\
  $^3$ Institute for Astronomy, ETH Z{\"u}rich, Wolfgang-Pauli-Strasse 27, CH-8093, Z{\"u}rich, Switzerland}

\begin{document}

\date{\today}
                             
\pagerange{\pageref{firstpage}--\pageref{lastpage}} \pubyear{2017}

\maketitle

\label{firstpage}

\begin{abstract}
The presence of an unseen `Planet~9' on the outskirts of the Solar system has been invoked to explain the unexpected clustering of the orbits of several Edgeworth--Kuiper Belt Objects. We use $N$-body simulations to investigate the probability that Planet~9 was a free-floating planet (FFLOP) that was captured by the Sun in its birth star-formation environment. We find that only 1 -- 6\,per cent of FFLOPs are ensnared by stars, even with the most optimal initial conditions for capture in star-forming regions (one FFLOP per star, and highly correlated stellar velocities to facilitate capture). Depending on the initial conditions of the star-forming regions, only 5 -- 10 of 10\,000 planets are captured onto orbits that lie within the constraints for Planet~9. When we apply an additional environmental constraint for Solar system formation -- namely the injection of short-lived radioisotopes into the Sun's protoplanetary disc from supernovae -- we find that the probability for the capture of Planet~9 to be almost zero.
\end{abstract}


\section{Introduction}

One of the outstanding issues in astrophysics is to understand the processes involved in planet formation and to characterise the Solar system within the context of other planetary systems. 
A recent body of research \citep[e.g.][]{Trujillo14,Batygin16a,Batygin16b,Brown16b,Holman16a,Holman16b} has suggested that the unusual orbital characteristics of several Edgeworth--Kuiper Belt Objects could be explained by the presence of an unseen/undiscovered $\sim$20\,M$_\oplus$ planet -- the so-called `Planet~9' \citep[though see][for arguments against this hypothesis]{Nesvorny17,Shankman17}. Simulations constrain Planet~9's semimajor axis to between 380 -- 980\,au, its perihelion to between 150 -- 350\,au, eccentricity between 0.34 -- 0.72 and its inclination to be less than 30 -- 60$^\circ$  \citep{Batygin16a,Batygin16b,Brown16b,Holman16a,Holman16b}.

The inherent difficulty of forming a relatively massive planet via core accretion at such a large distance from the Sun has fuelled speculation that -- if real -- Planet~9 may have been captured by the Sun \citep{Li16}, or even ``stolen'' from another star in the Sun's birth environment \citep{Mustill16}.

The Sun's current location in the Galaxy is a low-density stellar environment, where interactions with passing stars are rare. However, most stars are born in star-forming regions where the stellar density is much higher \citep[$1 - 10^6$ stars\,pc$^{-3}$,][]{Lada03,Porras03,Bressert10}. As planet formation occurs almost immediately after the onset of star formation \citep{Haisch01}, then the influence of neighbouring stars in the Sun's natal star-forming region cannot be neglected.

Furthermore, several authors have shown that stellar/sub-stellar objects can be readily captured during the evolution and dissolution of relatively dense ($>$100\,stars\,pc$^{-3}$) star-forming regions \citep{Moeckel10,Kouwenhoven10,Parker12a,Perets12}. In this paper we revisit this question and determine whether a significant fraction of stars can capture free-floating planetary-mass objects (FFLOPs) in their birth star-forming environment with orbital characteristics consistent with the hypothesised Planet~9. We describe our simulations in Section~\ref{methods}, we present our results and discussion in Section~\ref{results} and we conclude in Section~\ref{conclusions}.

\section{Method}
\label{methods}

We use $N$-body simulations to model the evolution of star-forming regions containing a population of free-floating planetary mass objects. Observations \citep{Elmegreen01,Cartwright04,Peretto06,Andre10,Henshaw17} and simulations \citep{Bonnell03,Dale12c,Girichidis12,Vazquez17} suggest that stars form in a spatially substructured distribution, with correlated velocities on local scales \citep{Larson81,Kauffmann13,Hacar13}. In order to mimic this in our simulations, we use fractal distributions as our initial conditions, using the method described in  \citet{Goodwin04a} to determine both the spatial and kinematic properties of our star-forming regions.

In each simulation, the fractal dimension is set to be $D = 1.6$, which corresponds to a high degree of spatial and kinematic substructure in three dimensions. Previous work \citep{Kouwenhoven10,Perets12} has shown that spatial and kinematic substructure in star-forming regions facilitates the capture of low-mass companions and that these objects are more likely to be captured if the amount of substructure is highest.

We then scale the velocities of the objects to a virial ratio, $\alpha_{\rm vir}$ ($\alpha_{\rm vir} = T/|\Omega|$, where $T$ and $|\Omega|$ are the total kinetic energy and total potential energy of the objects, respectively) so the star-forming regions undergo three different phases of bulk motion. Regions with $\alpha_{\rm vir} = 0.3$ have subvirial (cool) velocities and undergo collapse to form a star cluster, whereas regions with $\alpha_{\rm vir} = 0.7$ are mildly supervirial and gently expand. Finally, a third set of regions have $\alpha_{\rm vir} = 1.5$ and are unbound, undergoing rapid expansion. We refer the interested reader to \citet{Goodwin04a} and \citet{Parker14b} for full details of the set-up of these fractal initial conditions and examples of their dynamical evolution.

In our fiducial simulations, we draw 1000 single stars from a \citet{Maschberger13} initial mass function (IMF) with stellar mass limits between 0.1 -- 50\,M$_\odot$. We then add a further population of planetary-mass objects which is equal to the number of stars (i.e. a star-planet ratio of 1:1), apart from one set of simulations where we impose a star-planet ratio of 5:1. The planetary mass objects are all assigned the same mass, which is either Jupiter-mass (1\,M$_{\jupiter} = 1 \times 10^{-4}$M$_\odot$), or ten Earth-masses (10\,M$_{\oplus} = 3 \times 10^{-5}$M$_\odot$). Whilst Planet~9's mass has been constrained to $\sim$20\,M$_{\oplus}$, we include the more massive planets to highlight the very slight dependence of the results on planet mass, and to enable a comparison with previous work \citep{Parker12a,Perets12}. We vary the total number of stars (either 150 or 1000) and the initial radius of the star-forming regions (1 or 3\,pc), commensurate with observations of nearby star-forming regions \citep[e.g.][]{Lada03,Pfalzner16}. These values lead to initial densities that are much higher than the \emph{present day} densities in nearby star-forming regions \citep{Bressert10}, but these regions (and the Solar system's birth environment) may have been more dense initially. 

In our initial conditions, we do not make a direct assumption on the origin of the free-floating planets. However, because they are assigned positions and velocities in the same way as the stars, the implicit assumption is that they form ``like stars''. That said, planetary mass objects that have been liberated from host stars in simulations of dense star-forming regions often have the same spatial and kinematic distributions as stars \citep[e.g.][]{Parker12a}, so the FFLOPs in our simulations could in principle have two different origins. However we note that FFLOPs produced in planet--planet scattering events can have a different kinematic distribution to those liberated by interactions with passing stars.

We note that the expected number of FFLOPs in the Milky Way is uncertain, with some authors claiming one FFLOP per main sequence star \citep{Sumi11} \citep[although this result has been called into question by][]{Raymond11,Quanz12,Mroz17}. It is also difficult to pinpoint their origin (either they are an extension of the stellar mass function, or they are planets liberated from orbit around host stars). However, our simulations are designed to deliberately facilitate the capture of FFLOPS and we therefore create a large reservoir of these objects in our simulated star-forming regions.

We evolve our star-forming regions for 10\,Myr using the \texttt{kira} integrator in the \texttt{Starlab} environment \citep{Zwart99,Zwart01} with stellar evolution switched on using the \texttt{SeBa} package \citep{Zwart96,Zwart12}, also within \texttt{Starlab}. A summary of the different initial conditions is given in Table~\ref{initials}.

\begin{table}
  \caption[bf]{Summary of the initial conditions.  The columns show the number of stars, $N_{\rm stars}$, number of planets, $N_{\rm planets}$, the mass of the planet, $m_p$, the virial ratio,  $\alpha_{\rm vir}$, the radius of the star-forming region, $r_F$ and the median initial local density this mass and radius results in, $\tilde{\rho}_{\rm ini}$.}
  \begin{center}
    \begin{tabular}{|c|c|c|c|c|c|}
      \hline
      $N_{\rm stars}$ & $N_{\rm planets}$ & $m_p$ & $\alpha_{\rm vir}$ & $r_F$ & $\tilde{\rho}_{\rm ini}$  \\
      \hline
      1000 & 1000 & 1\,$M_{\jupiter}$ & 0.3 & 1\,pc & 5000 -- 30\,000\,M$_\odot$\,pc$^{-3}$ \\
      1000 & 1000 & 1\,$M_{\jupiter}$ & 0.7 & 1\,pc & 5000 -- 30\,000\,M$_\odot$\,pc$^{-3}$ \\
      1000 & 1000 & 1\,$M_{\jupiter}$ & 1.5 & 1\,pc & 5000 -- 30\,000\,M$_\odot$\,pc$^{-3}$ \\
      \hline
      1000 & 1000 & 10\,$M_{\oplus}$ & 0.3 & 1\,pc & 4000 -- 20\,000\,M$_\odot$\,pc$^{-3}$ \\
      1000 & 1000 & 10\,$M_{\oplus}$ & 0.7 & 1\,pc & 4000 -- 20\,000\,M$_\odot$\,pc$^{-3}$ \\
      1000 & 1000 & 10\,$M_{\oplus}$ & 1.5 & 1\,pc & 4000 -- 20\,000\,M$_\odot$\,pc$^{-3}$ \\
      \hline
      1000 & 1000 & 1\,$M_{\jupiter}$ & 0.7 & 3\,pc & 200 -- 700\,M$_\odot$\,pc$^{-3}$ \\
      \hline    
      1000 & 200 & 1\,$M_{\jupiter}$ & 0.3 & 1\,pc & 3000 -- 15\,000\,M$_\odot$\,pc$^{-3}$ \\
      1000 & 200 & 1\,$M_{\jupiter}$ & 0.7 & 1\,pc & 3000 -- 15\,000\,M$_\odot$\,pc$^{-3}$ \\
      \hline
      150 & 150 & 1\,$M_{\jupiter}$ & 0.7 & 1\,pc & 100 -- 800\,M$_\odot$\,pc$^{-3}$ \\
      150 & 150 & 1\,$M_{\jupiter}$ & 1.5 & 1\,pc & 100 -- 800\,M$_\odot$\,pc$^{-3}$ \\
      \hline
    \end{tabular}
  \end{center}
  \label{initials}
\end{table}

\section{Results}
\label{results}

\subsection{Fraction of captured planets}

\begin{figure}
  \begin{center}
    \rotatebox{270}{\includegraphics[scale=0.38]{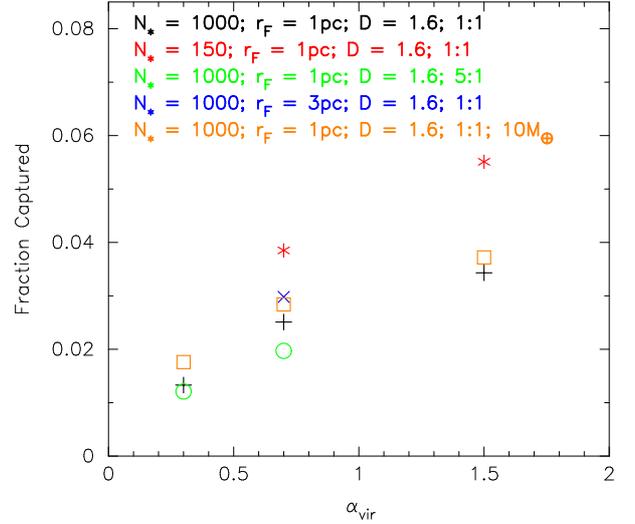}}
    \caption[bf]{The fraction of free-floating planets (FFLOPs) captured as a function of the initial dynamical state of the star-forming region. $\alpha_{\rm vir} = 0.3$ indicates a collapsing (bound, cool) region,  $\alpha_{\rm vir} = 0.7$ is a mildly expanding (supervirial, warm) region and $\alpha_{\rm vir} = 1.5$ is a rapidly expanding (unbound, hot) region. The coloured points indicate different initial conditions;  $N_\star$ is the number of stars, $r_F$ is the radius of the star-forming region, $D$ is the fractal dimension (amount of substructure) and the ratio of stars to FFLOPs. The FFLOPs are all Jupiter-mass, apart from the simulations shown by the orange points (10\,M$_\oplus$).}
      \label{capture_fraction}
  \end{center}
\end{figure}

Planets are captured by stars in all three sets of initial conditions for the evolution of our star-forming regions. In Fig.~\ref{capture_fraction} we show the fraction of captured planets, $f_{\rm cap}$ (the number in a bound orbit around a star, $N_{p, \rm bound}$, divided by the total number of planets in the simulation initially $N_{\rm planets}$) as a function of the initial virial ratio, or bulk motion, of the star-forming region. We have summed together ten realisations of each initial condition, identical apart from the random number seed used to set the positions and velocities of all objects, and the stellar masses.

We find a clear dependence of the fraction of captured planets on the initial virial ratio of the star-forming region. Regions that are subvirial ($\alpha_{\rm vir} = 0.3$), bound and subsequently collapse capture fewer planets than those that are very supervirial ($\alpha_{\rm vir} = 1.5$), unbound and rapidly expanding. This is due to a combination of two effects. Firstly, the unbound supervirial regions never fully dynamically mix, which enables the retention of kinematic substructure and the conditions to facilitate further capture of planets \citep[see also][]{Kouwenhoven10,Perets12}. The supervirial regions also expand rapidly, lowering the stellar density and preventing planets captured on fragile orbits with low binding energy from being disrupted by interactions with passing stars \citep{Parker14b}.

The fraction of captured FFLOPs strongly depends on the number of objects in the region initially, with regions with only $N_\star = 150$ (the red points in Fig.~\ref{capture_fraction}) capturing many more planets than our fiducial models ($N_\star = 1000$, $r_F = 1$pc, one FFLOP per star -- the black or orange points). This is not simply a density dependence, where lower-density regions can capture and retain more planets on fragile orbits; the blue point shows the capture fraction for an $N_\star = 1000$ region with a comparable stellar density to the $N_\star = 150$ regions  ($\sim$100\,stars\,pc$^{-3}$). Instead, the number of captured systems on wide and fragile orbits has been shown to be independent of the mass of a star-forming region \citep{Moeckel11b}, and so relatively more planets are captured in lower-$N$ star-forming regions of comparable density  to higher-$N$ regions.


The fraction of captured FFLOPs is not strongly dependent on the mass of the planets (compare the orange points for simulations with 10\,M$_\oplus$ planets to the black points for 1\,M$_{\jupiter}$ planets in Fig.~\ref{capture_fraction}), with lower-mass planets being slightly more susceptible to capture.
\vspace{-0.5cm}

\subsection{Orbital properties of captured planets}

\citet{Brown16b} show that Planet~9 is unlikely to be more massive than 20\,M$_\oplus$ and we therefore focus on our simulations where the FFLOPs are 10\,M$_\oplus$, although the following results do not depend on the FFLOP mass.

\begin{figure*}
  \begin{center}
\setlength{\subfigcapskip}{10pt}
\hspace*{-1.cm} 
\subfigure[Perihelia (dashed lines)/Semimajor axes (solid lines)]{\label{fflop_orbits-a}\rotatebox{270}{\includegraphics[scale=0.29]{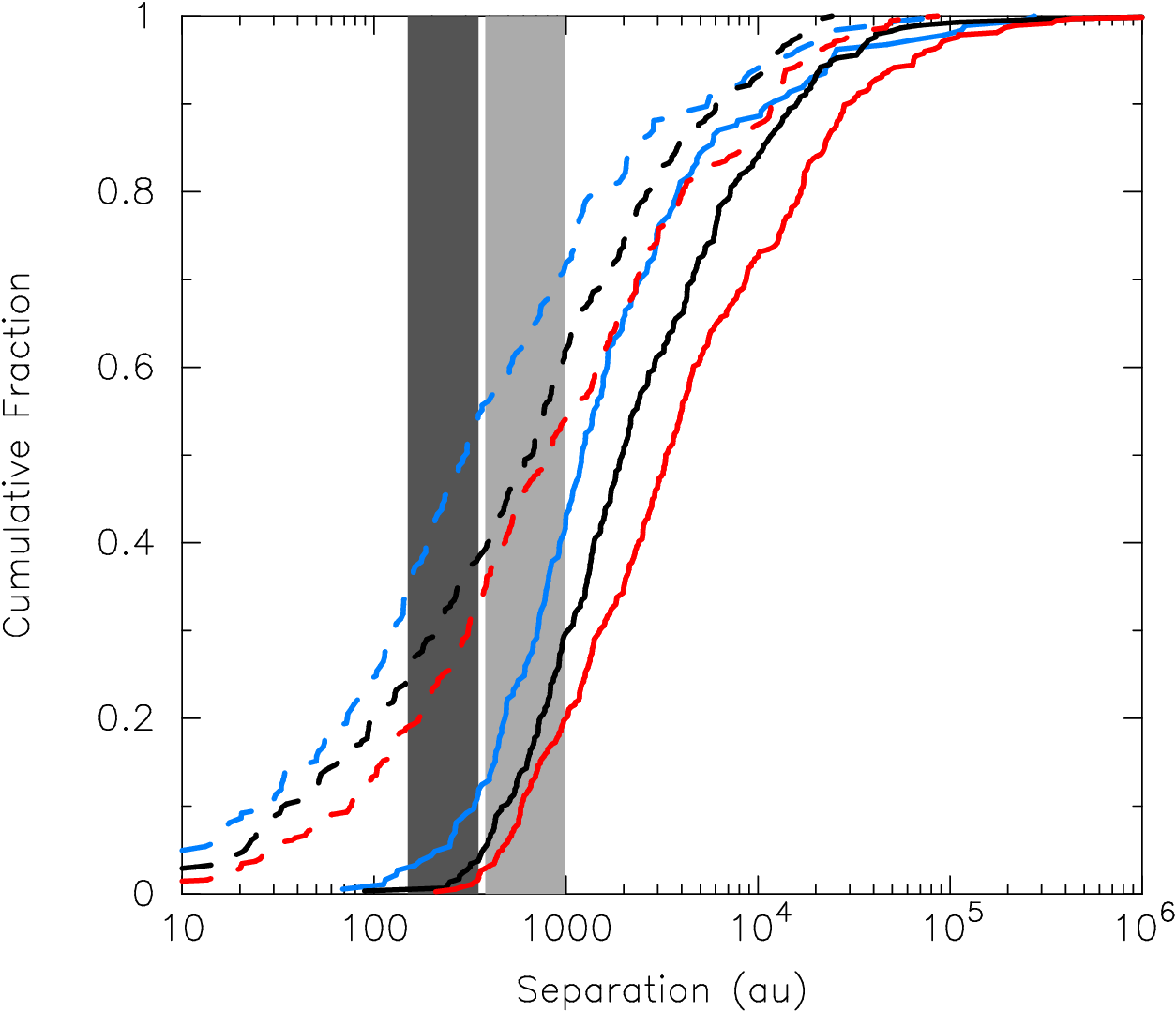}}}
\subfigure[Eccentricities]{\label{fflop_orbits-b}\rotatebox{270}{\includegraphics[scale=0.29]{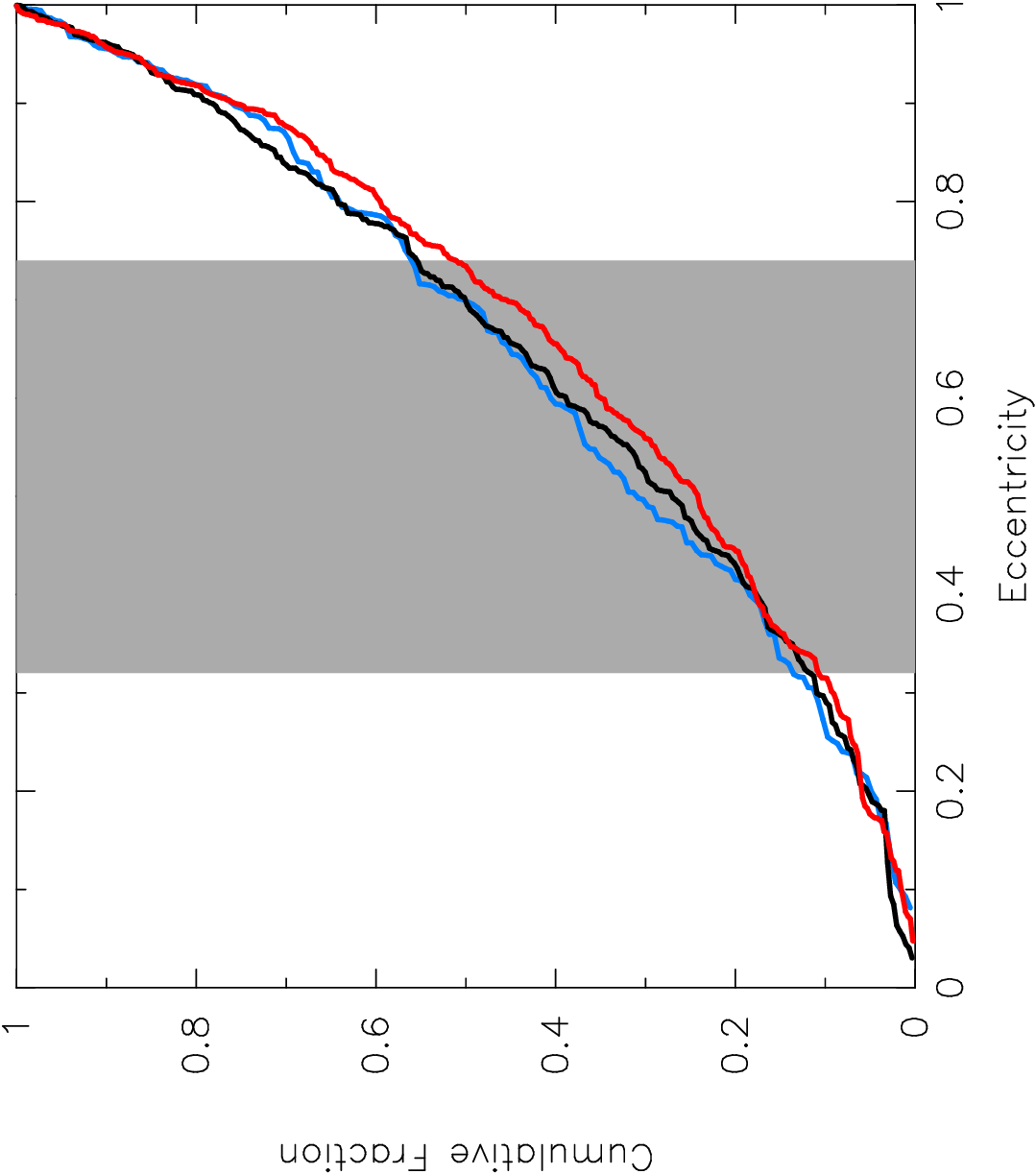}}}
\hspace*{0.1cm} 
\subfigure[Inclinations]{\label{fflop_orbits-c}\rotatebox{270}{\includegraphics[scale=0.29]{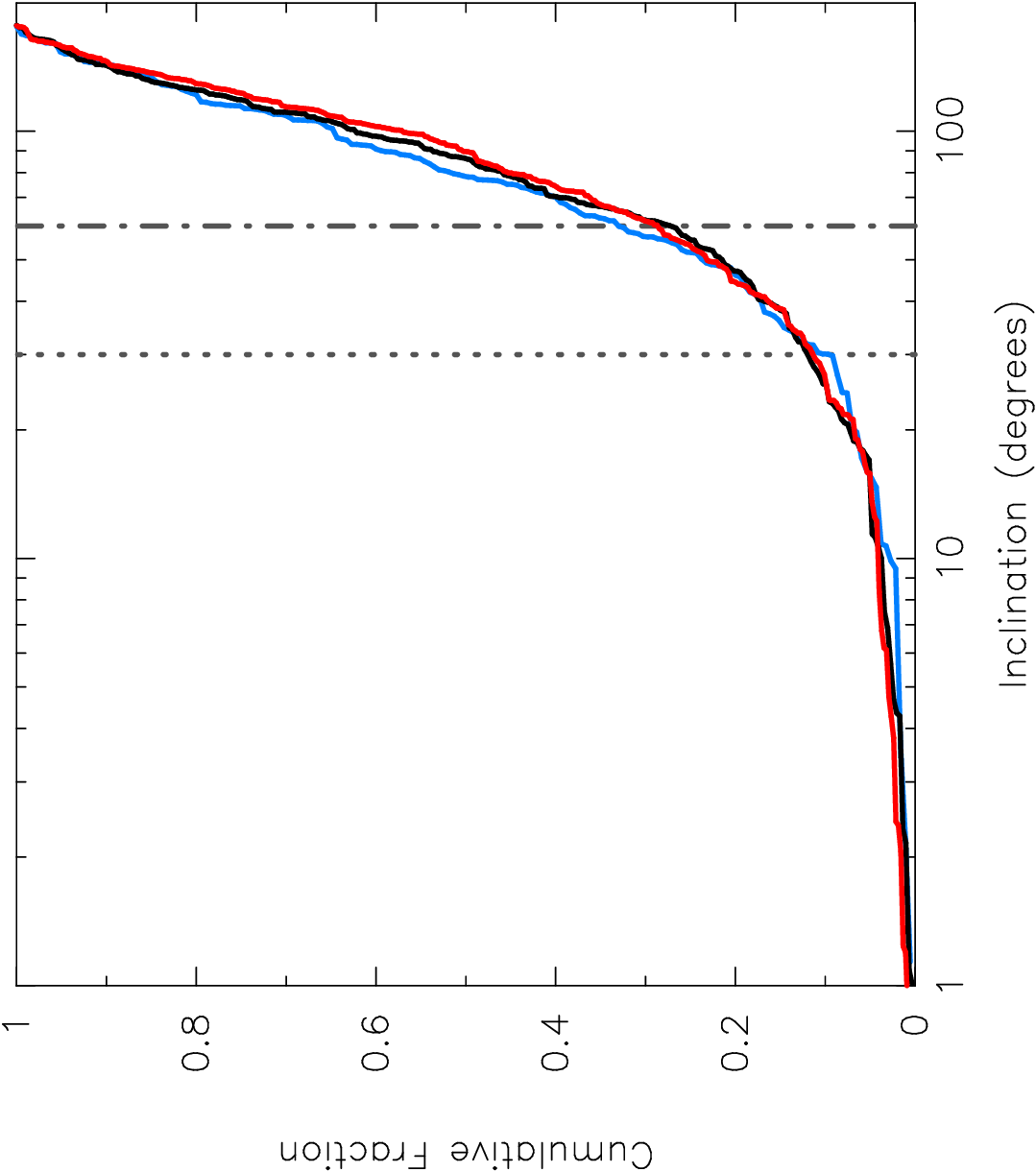}}} 
\caption[bf]{Orbital properties of captured FFLOPs in subvirial (collapsing) star-forming regions (blue lines), slightly supervirial (gently expanding) star-forming regions (black lines) and highly supervirial (unbound, rapidly expanding) star-forming regions (red lines). In panel (a) we show the perihelia of the captured planets by the dashed lines, and their semimajor axes by the solid lines. The shaded regions in panel (a) are limits on perihelion (dark grey) and semimajor axis (light grey) for Planet~9 from \citet{Brown16b}. In panel (b) we show the allowed range of eccentricities for  Planet~9s and in panel (c) the vertical lines are limits on inclination from \citet{Brown16b} and \citet{Holman16a,Holman16b}. Constraints on the orbit of Planet 9 favour inclinations lower than 30$^\circ$, but higher values ($< 60^\circ$) are not excluded.}
\label{fflop_orbits}
  \end{center}
\end{figure*}

In Fig.~\ref{fflop_orbits} we show the orbital parameters for the planets that are captured around a stellar-mass object in our simulations\footnote{We do not consider planet--planet systems \citep[which can form in this type of simulation, see also][]{Perets12} because we wish to examine systems where the primary-mass object is a star.}. In each panel in this figure, the cumulative distributions are shown for all three virial ratios in the simulations. 

The subvirial (collapsing) star-forming regions are more likely to capture a planet on an orbit with a perihelion distance $r_{\rm peri}$ in the range of allowed values for Planet~9 than regions with supervirial motion.  $r_{\rm peri}$ is defined in the usual way as
\begin{equation}
r_{\rm peri} = a_p\left(1 - e_p\right),
\end{equation}
where $a_p$ and $e_p$ are the semimajor axis and eccentricity of 
in the captured planet, respectively. The eccentricity distribution is roughly thermal (panel b), which is expected for any binary system that forms via capture \citep{Kroupa95a,Kouwenhoven10,Perets12}. Unlike the perihelion and semimajor axis, there is no dependence of the eccentricity or inclination (panel c) on the virial ratio of the star-forming region.

Whereas a planet is more likely to be captured by a star in an expanding, supervirial star-forming region, a captured planet is more likely to have the required orbital parameters for Planet~9 if it is captured in a subvirial, collapsing star-forming region (Fig.~\ref{fflop_orbits-a}).

However, the fraction of captured systems have orbital parameters in the range specified for Planet~9 is still extremely small. Summing together ten realisations of the same initial conditions for each simulation, we find that only $\sim$5--10 out of 10\,000 FFLOPs are captured with a perihelion in the range 150 -- 350\,au, semimajor axis between 380 -- 980\,au, eccentricity between 0.34 -- 0.72 and an inclination of less than 30 -- 60$^\circ$.
\vspace{-0.5cm}

\subsection{Planet 9 in the context of Solar system formation}

\begin{figure}
  \begin{center}
    \hspace*{-1.cm}
    \rotatebox{270}{\includegraphics[scale=0.38]{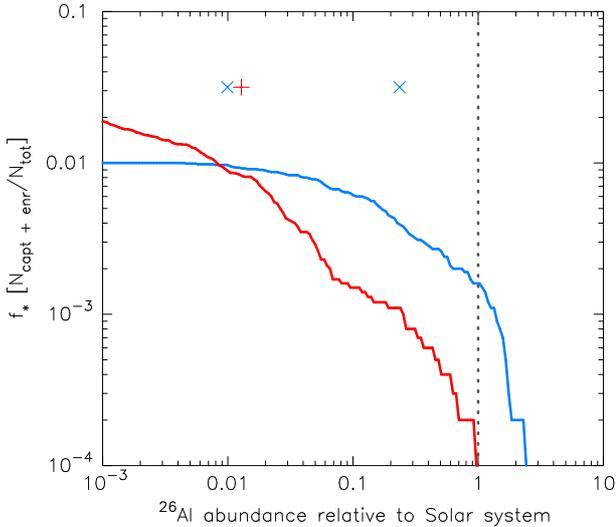}}
    \caption[bf]{Inverse cumulative distribution showing the estimated $^{26}$Al abundance relative to the Solar system value for stars in subvirial regions (blue line) and in highly supervirial regions (red line). The distribution is normalised to the fraction of stars that capture a planet \emph{and} are enriched by supernovae ejecta. We also show the $^{26}$Al abundance values for stars that capture a planet within the orbital constraints of Planet~9; the red plus sign is for a planet captured in the supervirial simulations, and the blue cross is for planets captured in the subvirial simulations (three in total). All other captured FFLOPs within the allowed Planet~9 parameter space have negligible  $^{26}$Al abundances.}
    \label{fflop_enrichment}
  \end{center}
\end{figure}

The possible capture of Planet~9 in the Sun's natal star-forming region is not the only line of argument that the Solar system formed in a dense stellar environment. Many authors have presented evidence that the Sun was either directly enriched by radiogenic isotopes \citep{Adams14,Lichtenberg16b,Parker16a,Nicholson17,Telus17} or formed from pre-enriched material \citep[e.g.][]{Cameron77,Gounelle15,Boss17} and both scenarios require the Sun to form in reasonably dense stellar environments \citep[$>$100\,stars\,pc$^{-3}$,][]{Parker16a}. 

Of these isotopes, the abundance of $^{26}$Al is the most robustly measured and can be used to estimate the amount of radiogenic heating during planet formation \citep[][and references therein]{Lichtenberg16a,Lichtenberg16b}, and to constrain the origin and dynamics of planet-forming material in the Solar system \citep[Lichtenberg et al., submitted;][]{Kita13}. Following the methodology in \citet{Lichtenberg16b}, in Fig.~\ref{fflop_enrichment} we show the inverse cumulative distribution of the  $^{26}$Al abundance compared to the Solar system value in either the subvirial (blue) or very supervirial (red) simulations. We show the canonical initial $^{26}$Al/$^{27}$Al ratio \citep{Kita13} for the Solar system by the vertical dashed line. The subvirial (collapsing) regions, which retain a higher stellar density throughout the simulations, are more conducive to isotope enrichment than supervirial (expanding) regions, which will likely be too diffuse at the time of the first supernovae \citep[4 -- 7\,Myr,][]{Parker14a,Parker14b}. The cumulative distribution is normalised to the fraction of stars that capture a FFLOP and are enriched (at a level $\geq 10^{-3}$ of the Solar value); this is a small percentage of the total number of stars, and only three of these stars capture a FFLOP on an orbit within the constraints of Planet~9 and experience enrichment anywhere near Solar system levels (the crosses/plus symbols in  Fig.~\ref{fflop_enrichment}). A further two stars are have  $^{26}$Al abundances less than $10^{-3}$ of the Solar value and are not shown. Finally, we note that the three stars that are enriched, and capture a planet, have stellar masses considerably lower than that of the Sun (0.78, 0.12 and 0.24\,M$_\odot$, in order of increasing $^{26}$Al abundance). 

\section{Conclusions}
\label{conclusions}

We present $N$-body simulations of the dynamical evolution of star-forming regions with a significant population of free-floating planetary mass objects (FFLOPs). We vary the initial virial ratio of the star-forming regions, so that they are either subvirial (bound) and collapse to form a star cluster, are slightly supervirial and gently expand, or are very supervirial (unbound) and rapidly expand. We vary the number of stars and the initial radii of the star-forming regions and we vary the mass of the FFLOPs, and the number per star in the star-forming regions. Our conclusions are the following:

(i) Between 1 and 6\,per cent of planets are captured by stars in star-forming regions with initial conditions optimised for the ensnarement of low-mass objects onto orbits around stars, when these regions contain a significant reservoir of FFLOPs available for capture.

(ii) The fraction of captured planets is a strong function of the initial virial ratio (bulk velocity) of the star-forming region, with FFLOPs twice as likely to be captured in supervirial (unbound) regions undergoing rapid expansion than in subvirial (bound) regions undergoing collapse.

(iii) However, planets captured in star-forming regions that collapse to form a cluster are more likely to have orbits consistent with the allowed parameter space of the proposed Planet~9.

(iv) Convolving the relative numbers of planets in (ii) and (iii), we find the number of planets fulfilling the orbital criteria for Planet~9 is then independent of the initial virial ratio of the star-forming region, and is extremely low (of order 5 -- 10 from an initial population of 10\,000 FFLOPs).

(v) Finally, we note that the Sun was likely enriched by a supervernova explosion in its birth environment. For this to occur, the initial conditions in the Sun's natal star-forming region are likely to have been subvirial in order to facilitate the collapse of the region to form a bound, relatively dense star cluster.


Overall, our results suggest that the fraction of stars that capture a FFLOP onto an orbit consistent with that of the hypothesised Planet~9, and that experience Solar system levels of isotope enrichment, is almost zero. This fraction is lower than that reported in \citet{Li16} and \citet{Mustill16}. This is likely due to differences in the assumed initial velocities of the stars (we adopt quasi-Gaussian kinematic substructure whereas the earlier work assumes a Maxwellian distribution), higher ($\times$10) initial densities (required to facilitate supernova enrichment), and our FFLOPs do not occur as the result of previous planet-planet scattering events. 
\vspace{-0.5cm}

\section*{Acknowledgements}

We thank the referee, Fred Adams, for his helpful comments and suggestions. RJP acknowledges support from the Royal Society in the form of a Dorothy Hodgkin Fellowship. TL was supported by ETH Research Grant ETH-17 13-1. Part of this work has been carried out within the framework of the National Center for Competence in Research ``PlanetS'', supported by the Swiss National Science Foundation. SPQ acknowledges the financial support of the SNSF.
\vspace{-0.5cm}

\bibliographystyle{mnras}
\bibliography{general_ref}

\label{lastpage}

\end{document}